\documentclass[usenatbib,referee]{mn2e}
\usepackage{amsmath}
\usepackage{graphicx}
\usepackage{epsfig}
\usepackage{txfonts}
\usepackage{color}
\usepackage{natbib}
\usepackage{harpoon}
\usepackage{multirow}

\title[Evolution of double ONe WD merger remnant]{Evolution of double oxygen-neon white dwarf merger remnant}
\author[C. Wu, H. Xiong, Z. Han, \& B. Wang]
{Chengyuan Wu$^{\rm 1,2,3}$\thanks{E-mail:wuchengyuan@ynao.ac.cn}, Heran Xiong$^{\rm 4}$, Zhanwen Han$^{\rm 1,2,3}$, Bo Wang$^{\rm 1,2,3}$\\
$^1$Yunnan Observatories, Chinese Academy of Sciences, Kunming 650216, China\\
$^2$Key Laboratory for the Structure and Evolution of Celestial Objects, Yunnan Observatories, CAS, Kunming 650216, China\\
$^3$International Centre of Supernovae, Yunnan Key Laboratory, Kunming, 650216, China\\
$^4$Research School of Astronomy and Astrophysics, The Australian National University, Canberra, ACT 2611, Australia\\}

\date{Accepted XXX. Received YYY; in original form ZZZ}

\pubyear{2023}

\begin{document}
\label{firstpage}
\pagerange{\pageref{firstpage}--\pageref{lastpage}}
\maketitle

\begin{abstract}

Double white dwarf (WD) merger process and their post-merger evolution are important in many fields of astronomy, such as supernovae, gamma-ray bursts, gravitational waves, etc. The evolutionary outcomes of double ultra-massive WD merger remnants are still a subject of debate, though the general consensus is that the merger remnant will collapse to form a neutron star. In this work, we investigate the evolution of a $2.20{M}_{\odot}$ merger remnant stemmed from the coalescence of double $1.10{M}_{\odot}$ ONe WDs. We find that the remnant ignites off-centre neon burning at the position near the surface of primary WD soon after the merger, resulting in the stable inwardly propagating oxygen/neon (O/Ne) flame. The final outcomes of the merger remnant are sensitive to the effect of convective boundary mixing. If the mixing cannot stall the O/Ne flame, the flame will reach the centre within $20$ years, leading to the formation of super Chandrasekhar mass silicon core, and its final fate probably be neutron star (NS) through iron-core-collapse supernova. In contrast, if the convective mixing is effective enough to prevent the O/Ne flame from reaching the centre, the merger remnant will undergo electron capture supernova to form an ONeFe WD. Meanwhile, we find that the wind mass loss process may hardly alter the final fate of the remnant due to its fast evolution. Our results imply that the coalescence of double ONe WDs can form short lived giant like object, but the final outcomes (NS or ONeFe WD) are influenced by the uncertain convective mixing in O/Ne flame.

\end{abstract}


\section{Introduction}

White dwarfs (WDs) are the most numerous members of the stellar graveyard. It is accepted that more than $95\%$ of all stars in the Universe will evolve into WDs (e.g. C{\'o}rsico et al. 2019). Double white dwarfs (WDs) in binary systems with sufficiently short orbital periods may merge within the Hubble time due to the angular momentum loss from the gravitational wave radiation. The coalescences of double white dwarfs are thought to be one of the most common evolutionary endpoints of binary systems, and may relevant to some important objects or astrophysical phenomena. For example, (1) type Ia supernovae may be stemmed from the mergers of double carbon-oxygen (CO) WDs or the mergers of CO WD and helium (He) WD (e.g. Iben \& Tutukov 1984; Webbink 1984; Han 1998; Fink et al. 2010; Wang 2018; Shen et al. 2018); (2) massive double WD mergers, containing at least one oxygen-neon (ONe) WDs are thought to be related to some accretion-induced-collapse (AIC) events or high-energy phenomena such as gamma-ray bursts or high-energy neutrino emissions (e.g. Lyutikov \& Toonen 2017; Ruiter et al. 2019); (3) double WD systems are verified gravitational wave (GW) sources (e.g. Kilic et al. 2014), which can be observed by the future space-based GW detectors like LISA or TianQin (e.g. Ruiter et al. 2010; Kremer et al. 2017; Wang \& Liu 2020; Huang et al. 2020).

Double CO or ONe WD mergers may produce massive remnants with mass greater than Chandrasekhar mass limit (${M}_{\rm {ch}}\approx1.38{M}_{\odot}$). The evolutionary fates of these massive remnants is still a subject of debate. For example, the coalescence of double CO WD is thought to be the progenitor of SN Ia, however, it was quickly pointed out that the rapid mass transfer in such merger process would lead to the off-centre carbon ignition, triggering the inwardly propagating carbon flame which may transform the remnant to an ONe composition (e.g. Nomoto \& Iben 1985; Saio \& Nomoto 1985; Kawai et al. 1987; Chen et al. 2014). For the more massive double ONe WDs, although the general consensus is that the merger remnant will collapse to form a neutron star (NS) (e.g. Miyaji et al. 1980; Schwab et al. 2015; Wu \& Wang 2018), recent studies claimed that the uncertainties of the outcomes of Chandrasekhar mass ONe WD may arise from the competition between the thermal nuclear runaway of O/Ne and electron capture reactions of magnesium (Mg) and Ne. If an ONe WD can undergo thermal explosion, its light curve may similar to that of 1991T-like events (e.g. Marquardt et al. 2015). On the contrary, if the central density of ONe WD is able to achieve the order of $\sim{10}^{10}\,{\rm g}\,{\rm {cm}^{-3}}$, electron-capture reactions on $^{\rm 20}{\rm Ne}$ and $^{\rm 24}{\rm Mg}$ likely cause the WD to collapse into NS (e.g. Jones et al. 2016, 2019).

A theoretical method in understanding the final fates of double WD merger remnants is to construct the model of remnant that mimics the results of the SPH simulations using 1D stellar evolutionary code and then follow its secular evolution. Although the evolution of double CO WD and CO+He WD merger remnants have been widely studied so far (e.g. Yoon et al. 2007; Zhang et al. 2014; Schwab et al. 2016; Brooks et al. 2017; Wu et al. 2019; Wu \& Wang 2019; Lauer et al. 2019; Schwab 2021; Wu et al. 2022), there are only few works pay attention to the evolution of ultra-massive double WD mergers in which the remnants are composed of degenerate ONe core. Brooks et al. (2017) investigated the evolution of ONe WD+He WD, and argued that the merger remnant will evolve to He-giant and finally experience supernova explosion to form a fast and luminous transient. Kashyap et al. (2018) carried out the first fully 3D simulation of the merger of a $1.20{M}_{\odot}$ ONe WD with a $1.10{M}_{\odot}$ CO WD, and found that material from the CO WD secondary may readily to detonation during the merger, resulting in the faint and rapidly fading transient. More recently, Wu et al. (2023) investigated the evolution of a series of ONe WD+CO WD merger remnants. They found that if the remnants are more massive than a critical mass which is between $1.90{M}_{\odot}$ to $1.95{M}_{\odot}$, off-central Ne ignition will occur soon after the merger, resulting in the formation of ONeFe WDs through electron-capture supernova (ECSN) explosions. While, if the remnants are less massive than the critical mass, the final outcome of the remnants would be either NSs through ECSNe or ONe WDs, which is depended on the wind mass-loss process during the evolution. 

The evolution of double ONe WD merger remnants have not been well studied so far. Many questions such as whether Ne detonation would occur during merger process, how fast is the inwardly propagating Ne flame, and what is their evolutionary outcome are still remain uncertain. In this work, we construct the 1D structure of double $1.10{M}_{\odot}$ ONe WD merger remnant based on the information from Dan et al. (2014), and to follow its post-merger evolution. The article is organized as follows. In Sect.\,2, we introduce our initial models of merger remnant. The input physics and the results of simulations are shown in Sect.\,3. We discuss the model uncertainties in Sect.\,4, and summarize our results in Sect.\,5.

\section{Initial models} \label{sec:models}

In a double WD system, the orbital shrink due to the gravitational wave radiation could cause the secondary (less massive WD) fill in its Roche Lobe (RL) and transfer material to the primary WD. Previous works indicated that the mass ratio of two WDs, $q={m}_{2}/{m}_{1}$ (${m}_{1}$ and ${m}_{2}$ are the masses of primary and secondary WDs, respectively), is a pivotal parameter in affecting the mass transfer process. If $q$ is larger than the critical mass ratio,
\begin{equation}
    {q}_{\rm {crit}}=\frac{5}{6}+\frac{1}{2}\frac{{\rm {dln}R}}{{\rm {dln}}M},
\end{equation}
where $R$ and $M$ are radius and mass of the secondary, the increase of the radius of the secondary due to the reduction of its mass will exceed the increase of its RL radius caused by the transfer of orbital angular momentum, resulting in the dynamically unstable transfer and then leads to the coalescence of two WDs (e.g. Pringle \& Webbink 1975; Tutukov \& Yungelson 1979). The criterion of stable or unstable mass transfer of two WDs is influenced by the synchronization time-scale of spin-orbit coupling, ${\tau}_{\rm s}$. Under strong coupling (${\tau}_{\rm s}\rightarrow{0}$), ${q}_{\rm {crit}}\approx\frac{2}{3}$. By contrast, if ${\tau}_{\rm s}\rightarrow{\infty}$, i.e. no angular momentum is transferred from the primary to the secondary, ${q}_{\rm {crit}}$ could decrease to about $0.2$ (e.g. Marsh et al. 2004).

Smoothed particle hydrodynamics (SPH) simulations show that the unstable mass transfer process could lead to the merger of double WDs. The primary WD will be survived during the merger, whereas the secondary will be tidally disrupted in which a significant amount of material will form a rotationally supported disc surround the primary WD (e.g. Guerrero et al. 2004). In order to understand the conditions of hydrodynamical burning minutes after double WD merger, Dan et al. (2014) performed a large grid of double WD merger systems, covering a wide range of WD masses and chemical compositions. They found that the merger remnants are divided into four components in their simulations: cool degenerate core, hot envelope, Keplerian disc and tidal tail. They provided 1D approximate formulas for the masses of these regions. The fitting formulas for the mass coordinate of highest temperature in the envelope, $M$(${T}_{\rm {max}}$), and the masses of each components are as follows:
\begin{equation}
    {M}({T}_{\rm {max}})={M}_{\rm {tot}}(0.863-0.3335q),
\end{equation}
\begin{equation}
    {M}_{\rm {core}}={M}_{\rm {tot}}(0.7786-0.5114q),
\end{equation}
\begin{equation}
    {M}_{\rm {env}}={M}_{\rm {tot}}(0.2779-0.464q+0.716{q}^{2}),
\end{equation}
\begin{equation}
    {M}_{\rm {disc}}={M}_{\rm {tot}}(-0.1185+0.9763q-0.6559{q}^{2}),
\end{equation}
where, ${M}_{\rm {core}}$, ${M}_{\rm {env}}$, ${M}_{\rm {disc}}$, ${M}_{\rm {tot}}$ and $q$ are the mass of cool degenerate core, hot envelope, Keplerian disc, total mass of the merger remnant and the mass ratio of double WDs before merger, respectively. We use their provided fitting formulae to construct the merger structure of double ONe WDs and investigate its further evolution. Since the tidal tail accounts for only a few percent of the total mass, same as previous works, we do not consider the mass of tidal tail and that unbound from the system in our simulation (e.g. Schwab 2021; Wu et al. 2023). In this case, the initial remnant mass remains the same as the total mass of the merging WDs. 

We construct the merger structure by employing stellar evolution code $\tt{MESA}$ (version: 12778) (e.g. Paxton et al. 2011; 2013; 2015; 2018; 2019). In our calculation, we consider an $2.20{M}_{\odot}$ remnant which is stemmed from the merger of double $1.10{M}_{\odot}$ ONe WDs. We first create a $2.20{M}_{\odot}$ helium-pre-main-sequence-star (equals to the total mass of double ONe WDs) and evolve it until central density (${\rho}_{\rm c}$) achieves ${10}^{5}\,{\rm g}\,{\rm {cm}^{-3}}$. During the evolution, we ignored all the nuclear reactions and mixing processes to maintain the abundance unchanged. At the end of the first step, the central temperature of the remnant reaches ${\rm {log}}({T}/{\rm K})=8.77$, and the corresponding ${\rho}-{T}$ structure can be represented by the blue dotted line in Fig.\,1. 

Secondly, we use the built-in ${\tt {MESA}}$ capability to adjust the elemental abundance distribution of the merger remnant to the desired mass fraction. According to Dan et al. (2014), the initial abundance of ONe WD is: $^{\rm 16}{\rm O}=60\%$, $^{\rm 20}{\rm {Ne}}=35\%$ and $^{\rm 24}{\rm {Mg}}=5\%$. In our simulation, we adjust the abundance of our model to the uniform distribution same as Dan et al. (2014).

Finally, we injected energy into different mass shell of the remnant to construct the thermal structure of the merger remnant. The prescription of energy injection is similar to that described in Schwab (2021). For $0\leq{M}_{\rm r}\leq{M}_{\rm {core}}$,
\begin{equation}
    T({M}_{\rm r})={T}_{\rm {core}}.
\end{equation}
For ${M}_{\rm {core}}\leq{M}_{\rm r}\leq{M}_{\rm {core}}+{M}_{\rm {env}}$,
\begin{equation}
    s({M}_{\rm r})={s}_{\rm {core}}+[{s}_{\rm {env}}-s({M}_{\rm {core}})]\frac{{M}_{\rm {r}}-{M}_{\rm {core}}}{{M}_{\rm {peak}}}.
\end{equation}
And for ${M}_{\rm {core}}+{M}_{\rm {env}}\le{M}_{\rm r}\leq{M}_{\rm {tot}}$,
\begin{equation}
    s({M}_{\rm r})={s}_{\rm {env}}.
\end{equation}
In these equations, ${M}_{\rm {core}}$, ${M}_{\rm {env}}$ and ${M}_{\rm {peak}}$ are core masses, envelope masses and the mass coordinates of the maximum temperature of the merger remnants, respectively. ${T}_{\rm {core}}$ is the temperature of isothermal core, which are set to be ${10}^{8}{\rm K}$. ${s}_{\rm {core}}$ and ${s}_{\rm {env}}$ are, the specific entropies of the core and the envelope, respectively. We set ${s}_{\rm {core}}$ equals to the specific entropy of the core when ${T}_{\rm {core}}={10}^{8}\,{\rm K}$, and ${s}_{\rm {env}}$ equals to the following equation:
\begin{equation}
    {\rm log}({s}_{\rm {env}}/{\rm {erg}}\,{\rm {g}^{-1}}\,{\rm {K}^{-1}})=8.7+0.3({M}_{\rm {tot}}/{M}_{\odot}-1.5).
\end{equation}

After the energy injection process, we obtained the initial 1D structure of the merger remnant. The ${\rho}-{T}$ profile is shown as the red solid line in Fig.\,1. The merger remnant has an $0.576{M}_{\odot}$ isothermal core, and the mass coordinate of ${T}_{\rm {max}}$ is at $1.091{M}_{\odot}$. By comparison, the density and temperature at the ${T}_{\rm {max}}$ position in our model is consisted with that in SPH simulations provided by Dan et al. (2014).

\begin{figure*}
\begin{center}
\epsfig{file=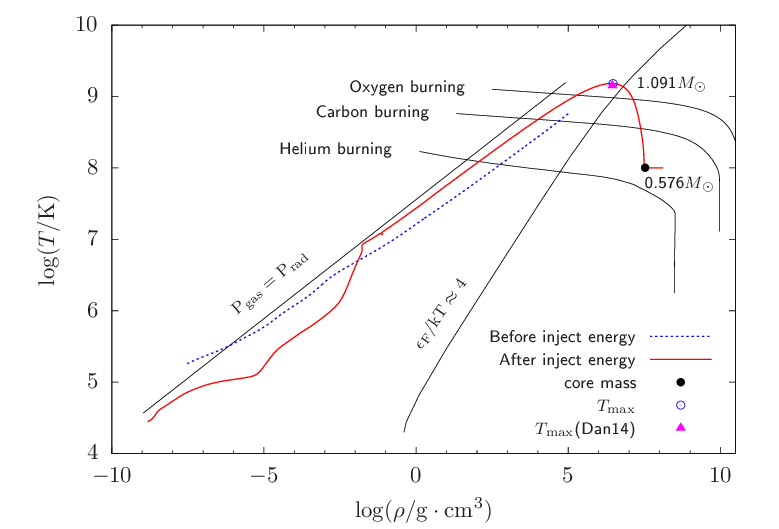,angle=0,width=10.2cm}
 \caption{Density-temperature (${\rho}-{T}$) structure diagram of the double $1.10{M}_{\odot}$ ONe WD merger remnant before (blue dashed line) and after (red solid line) energy injection. Black solid lines indicate the (${\rho}-{T}$) limits for electron degeneracy, radiation pressure equals to pressure of gas, He, C and O burning. Black filled circle and blue circle represent the position of the boundary of degenerate core and the ${T}_{\rm {max}}$ of our model, respectively. Masses of the corresponding positions are marked nearby. The magenta triangle represents the position of ${T}_{\rm {max}}$ given in Dan et al. (2014).}
  \end{center}
    \label{fig: 1}
\end{figure*}

\section{Evolution of the remnant} \label{sec:evolution}

\subsection{Input physics}

The composition of the merger remnant is oxygen enhancement. We use the OPAL radiative opacities for C/O-rich mixtures (e.g. Iglesias \& Rogers 1993, 1996), which are referred to as OPAL ``Type 2'' tables in ${\tt {MESA}}$. As the remnants evolve to giant phase, the surface temperature may decrease to lower than ${\rm log}(T/{\rm K})=3.70$. Since the lower boundary of the OPAL tabulations is ${\rm log}(T/{\rm K})=3.75$, and ${\tt {MESA}}$ does not provide low-temperature opacities that include separate C/O enhancements. Hence, similar to the prescription described in Schwab et al. (2016), we adopt Kasen opacities (${\rm {log}}({T}/{\rm K})\sim3.4-4.1$ to deal with the low temperature conditions, and blend the OPAL and Kasen opacities between ${\rm log}({T}/{\rm K})=4.1$ and $4.2$. This supplement is suitable for H- and He-deficient cool giant-like stars.

The nuclear reaction network adopted in the present work is ``${\tt {approx21.net}}$'' from ${\tt {MESA}}$ default, which is a 21-isotope $\alpha$-chain nuclear network. As the remnant expands toward giant phase, the superadiabatic gradient arising in radiation-dominated envelopes can extremely reduce computational efficiency. To circumvent this issue, we apply the treatment of convective energy transport ``MLT++'' (e.g. Paxton et al. 2013) to reduce the superadiabaticity in the radiation-dominated convective regions.  We constrain the time resolution in our calculations by adopting ``${\tt{varcontral\_target=1d-4}}$''. In order to have a better treatment on the propagation of off-center flame, we control the spatial resolution by changing ``${\tt {mesh\_dlog\_burn\_o\_dlogP\_extra=0.1}}$'', ``${\tt {mesh\_dlog\_burn\_ne\_dlogP\_extra=0.1}}$'' and ``${\tt {mesh\_dlog\_oo\_dlogP\_extra=0.1}}$'' to increase the mesh grid nearby the flame.

\subsection{Neon ignition and flame}

After obtained the profile of merger remnant, we restored the nuclear reactions to evolve the remnant forward in time. The remnant initially undergo the Kelvin–Helmholtz contract to increase the temperature of O/Ne shell. The contract process lasts for about $8.85$ hours when the temperature at the base of O/Ne shell achieves ${\rm {log}}({T}/{\rm K})\approx9.182$. Various properties and elemental abundance distribution of the merger remnant at the onset of Ne ignition are shown in Fig.\,2. Ne ignition occurs at the position very close to the surface of primary ONe WD where ${M}_{\rm r}\approx1.08{M}_{\odot}$ and ${R}\approx3000{\rm {km}}$. Due to the initially compact configuration of the merger remnant, the neon ignition begins in a nuclear burning flash. Afterwards, the envelope quickly inflate to giant structure where the energy produced by the nuclear reactions balance with the thermal energy transported from convection and neutrino loss. The evolution of the remnant on Hertzsprung-Russell (HR) diagram is shown in Fig.\,3.

\begin{figure*}
\begin{center}
\epsfig{file=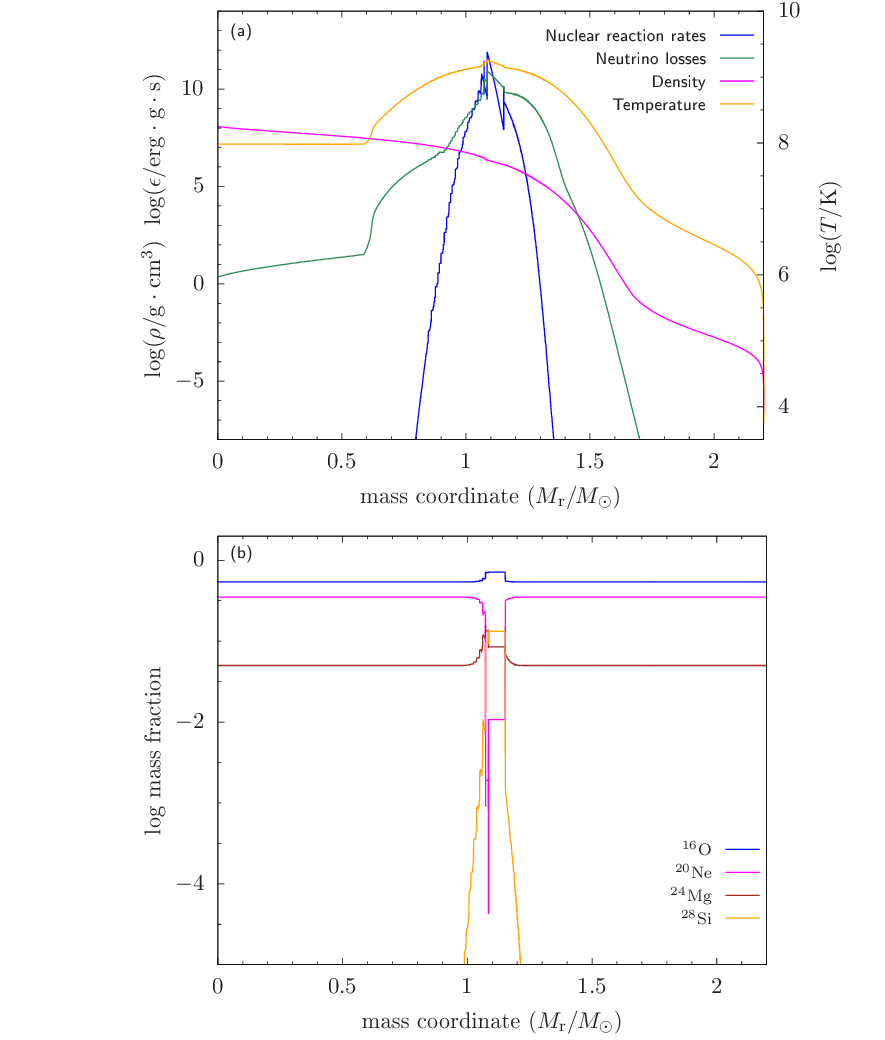,angle=0,width=10.2cm}
 \caption{Properties (panel a) and elements abundance distributions (panel b) of the $2.20{M}_{\odot}$ double ONe WD merger remnant at the onset of neon ignition. Blue, green, magenta and golden solid lines in panel (a) represent nuclear reaction rate of neon burning, energy loss rate due to neutrinos, density and temperature profiles, respectively. Mass fractions of four major elements ($^{\rm 16}{\rm O}$, $^{\rm 20}{\rm Ne}$, $^{\rm 24}{\rm Mg}$ and $^{\rm 28}{\rm Si}$) are shown in panel (b).}
  \end{center}
    \label{fig: 2}
\end{figure*}

At the onset of neon ignition, the temperature at the base of the shell is not high enough to lead to the substantial inwardly motion of the burning front by conduction. As the layer directly beneath the burning layer is heated up primarily via electron conduction from the burning layer above, the temperature of O/Ne shell increases until it is hot enough to sustain its own O/Ne burning. In this case, the steady O/Ne burning starts and quickly forms a O/Ne burning front which begins to propagate inwards towards the centre of the remnant. During this stage, a portion of energy of O/Ne burning used to heat the core and the others expands the envelope. The flame propagates inwards continuously, living behind burning ashes mainly consists of silicon-group (Si-group) elements. As can be seen from Fig.\,4, the location of the outer boundary of the convection zone varies significantly from timestep-to-timestep because of the high temperatures and energy generation rates associated with O/Ne fusion. The density of the material beneath the flame increases as the flame moves inward, making the timestep of calculation become extremely short. We stop our calculation when the flame almost reaches the centre (${M}_{\rm r}=2.38\times{10}^{-6}\,{M}_{\odot}$). At this moment, the remnant has evolved for about $20$ years, which costs approximately three wall-clock months and more than $9$ million timesteps of calculation. It takes about $20$ years for the flame to propagates $3000{\rm {km}}$, which means that the average velocity of the flame is about  $0.475{\rm {cm}}/{\rm s}$. For more rigorous estimate, Timmes et al. (1994) deduced that the velocity of off-centre burning flame is about 
\begin{equation}
{v}_{\rm flame}\approx(\frac{{\rm c}{\epsilon}_{\rm nuc}}{{\kappa}{\rho}{E}})^{\frac{1}{2}},
\end{equation}
and the flame width is about (e.g. Woosley \& Heger 2015)
\begin{equation}
{\delta}_{\rm flame}\approx(\frac{{\rm c}{E}}{{\kappa}{\rho}{\epsilon}_{\rm nuc}})^{\frac{1}{2}},
\end{equation}
where, $c$, ${\epsilon}_{\rm nuc}$, $\kappa$, $\rho$, and $E$(equals to ${C}_{\rm P}{T}$) are speed of light, nuclear burning rate, opacity, density and internal thermal energy per gram, respectively. We use the corresponding equation to estimate the theoretical velocity, which is shown as blue solid line in Fig.\,5. Initially, the flame moves relatively fast. As the flame moves inward, the density beneath the front of the flame increases, which slows down the propagation of the flame. The velocity decreases to about $0.2-0.5{\rm {cm}}/{\rm s}$ $570$ days after O/Ne ignition when the flame reached the mass coordinate of about $0.58{M}_{\odot}$. On the other hand, we can also estimate the velocity of neon flame from the numerical data. The Eulerian velocity of the flame can be determined by ${\rm d}{R}({T}_{\rm {max}})$/${\rm d}{t}$, where, ${R}({T}_{\rm {max}})$ is the radius at the maximum temperature (the position of the flame). However, the inner core undergoes thermal pulsation during the O/Ne burning, making the ${R}({T}_{\rm {max}})$ is difficult to refer to the position of the flame. We thus use the Lagrangian velocity to indicates the numerical flame velocity, i.e. 
\begin{equation}
{v}_{\rm flame}\approx{\dot{M}}/(4{\pi}{{r}^{2}}{\rho}),
\end{equation} 
Where, ${\dot{M}}$ is the mass of the shell that the flame crosses during per unit time. The Lagrangian velocity of the O/Ne flame is shown as red solid line in Fig.\,5. The shapes of the two curves match, which means that the numerical results are consistent with the theoretical results.

\begin{figure*}
\begin{center}
\epsfig{file=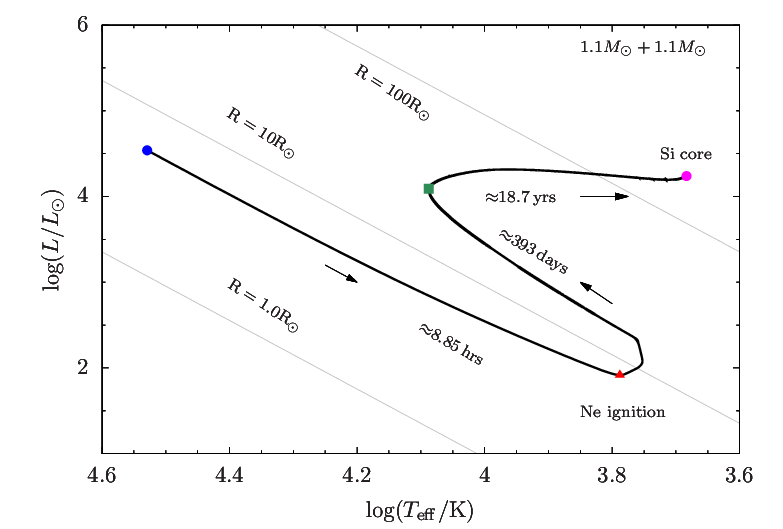,angle=0,width=10.2cm}
 \caption{The HR diagram of the merger remnant. The remnant undergoes Kelvin–Helmholtz contract and inflates to giant structure. Grey solid diagonal lines represent all of the locations where stars are found that have the same radius. Different evolutionary durations are shown in the diagram.}
  \end{center}
    \label{fig: 3}
\end{figure*}

\begin{figure*}
\begin{center}
\epsfig{file=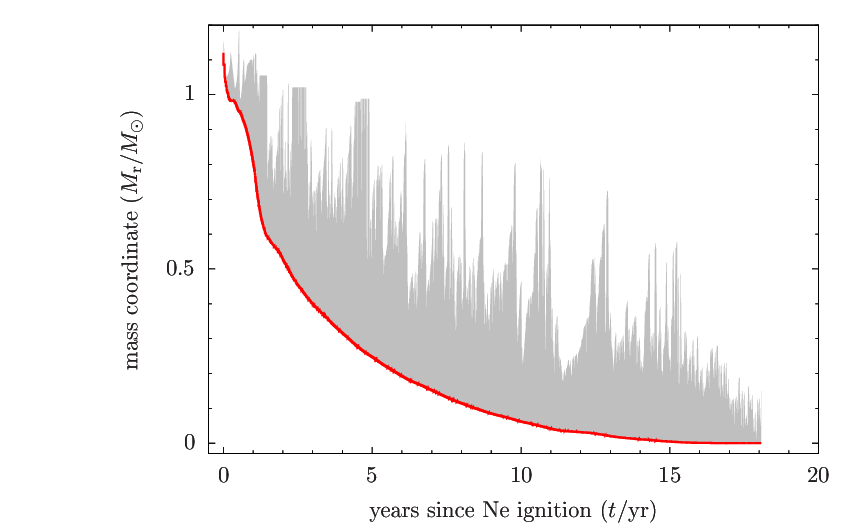,angle=0,width=10.2cm}
 \caption{The propagation of neon flame with no convective boundary mixing. x-axis shows the time since neon ignition whereas y-axis represents the Lagrangian mass coordinate. Shaded region represents the convective zone associated with flame, whereas the red solid line at the bottom of the region represents the position of the flame.}
  \end{center}
    \label{fig: 4}
\end{figure*}

\begin{figure*}
\begin{center}
\epsfig{file=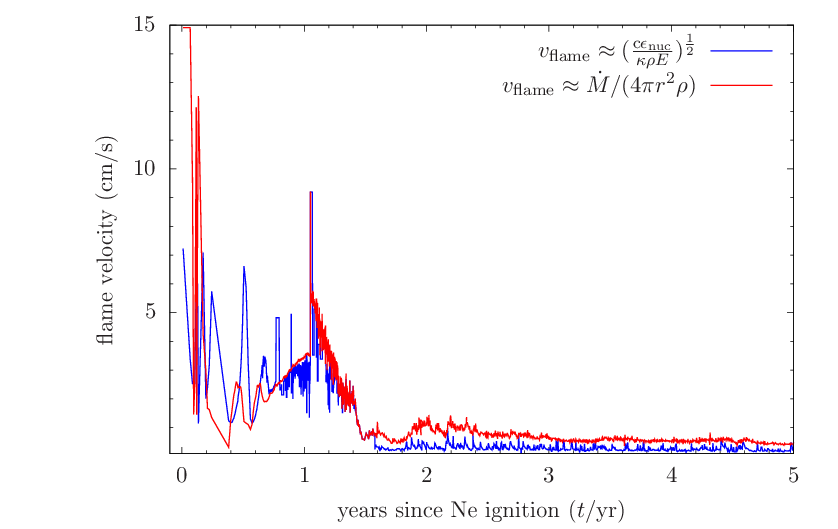,angle=0,width=10.2cm}
 \caption{The velocity of the inwardly propagating neon flame changing with time since neon ignition. Blue and red solid lines represent the theoretical ﬂame velocity and Lagrangian (numerical) velocity, respectively.}
  \end{center}
    \label{fig: 5}
\end{figure*}

\begin{figure*}
\begin{center}
\epsfig{file=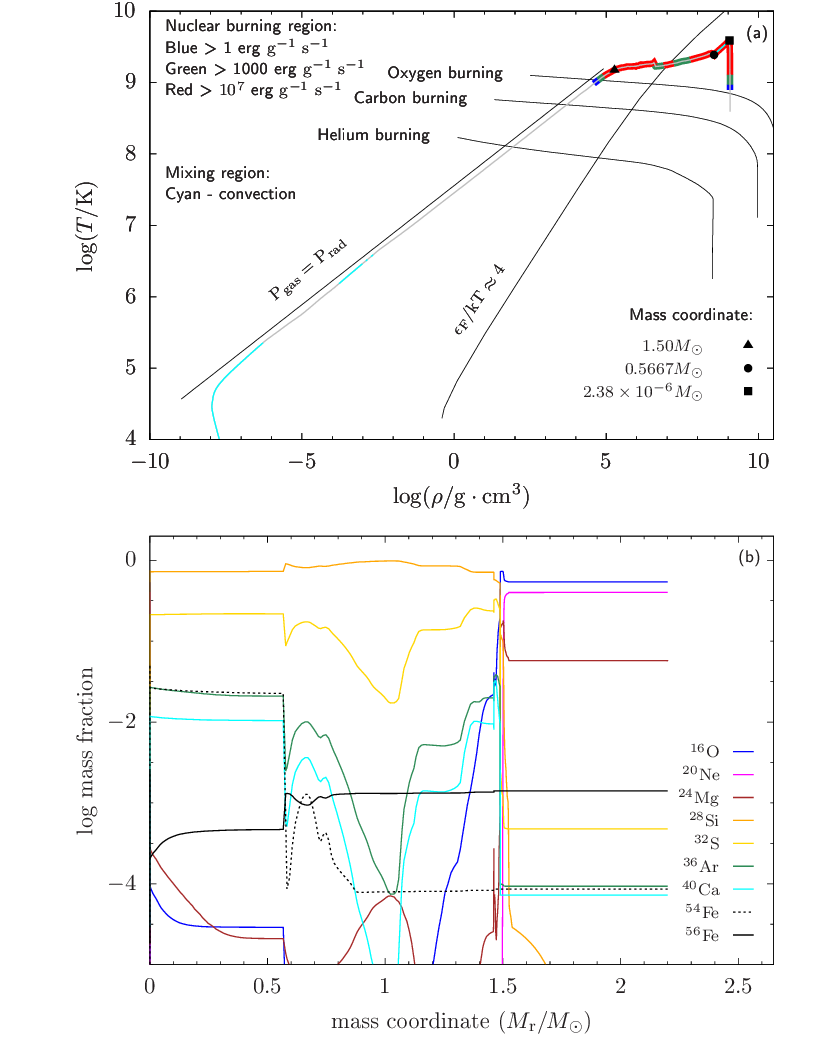,angle=0,width=10.2cm}
 \caption{Properties of the merger remnant when the neon flame moves close to the centre (${M}_{\rm r}=2.38\times{10}^{-6}\,{M}_{\odot}$). Panel (a): T-Rho profile, where black solid lines represent Pgas= Prad, degenerate line, burning line for helium, carbon, oxygen, respectively. Cyan line represents the convective zone, while blue, green and red thick lines represent burning region with different nuclear reaction rates, respectively. Triangle, cycle and square represent the positions of the boundary of core, central convective zone, and the ${T}_{\rm {max}}$, respectively. Panel (b): elements abundance distributions for different elements.}
  \end{center}
    \label{fig: 6}
\end{figure*}

\subsection{Final fates}

As the O/Ne flame migrates to the centre, it leaves behind hot Si-group ashes ($^{\rm 28}{\rm Si}$ and $^{\rm 32}{\rm S}$). As mentioned above, the timescale of calculation becomes extremely short when the neon flame moves close to the centre, and we do not follow the further evolution of merger remnant since it is very time consuming. We stop the simulation when the flame reached ${M}_{\rm r}=2.38\times{10}^{-6}\,{M}_{\odot}$. Fig.\,6 shows the ${\rho}-{T}$ and elemental abundance profiles of the merger remnant at the time when we stop the simulation. At this moment, the remnant is consisted of an $1.5{M}_{\odot}$ core with Si-group elements and an $0.7{M}_{\odot}$ ONeMg envelope. We predict that the flame will eventually reach the centre, and then undergo Kelvin-Helmholtz contract. Still, some Si flashes occur, leading to the off-centre Si burning. Schwab et al. (2016) found that the off-centre Si ignition occurs when the remnant mass is greater than $1.41{M}_{\odot}$. The mass coordinate of ${T}_{\rm {max}}$ in Si ashes of our model is at about $1.0{M}_{\odot}$, which means that the off-centre Si ignition may occur at this position. Subsequently, the Si flame will move towards the centre, and if the inwardly propagating Si flame could finally reach the centre, the super Chandrasekhar remnant will consist of an iron core, resulting in the formation of single NS through iron-core-collapse supernova.

If we assume that the outer ONeMg envelope could be ejected through the supernova explosion, the remnant will form an $1.5{M}_{\odot}$ NS with $0.7{M}_{\odot}$ ejecta. Owing to the low ejecta mass and the outer composition of the remnant, the supernova explosion may appear as a fast evolved subluminous transient, similar to some particular type Ic supernovae (e.g. Tauris et al. 2013, 2015 Moriya et al. 2017; Pellegrino et al. 2022). Note that for another possibility, if the collapsing core of the merger remnant could remain sufficient angular momentum to support an accretion disk or belt around the newly born NS, the jet launched from the disk collides with the circumstellar medium that was ejected during the merger of double WD. The interaction process could transform portion of kinetic energy of the jet into radiation, which may cause the formation of energetic supernova (e.g. Soker 2016, 2019, 2022; Gilkis \& Soker 2015; Gilkis et al. 2016; Kaplan \& Soker 2020).

\section{Discussion}

\subsection{Convective boundary mixing}

At the onset of evolution, the merger remnant undergoes the Kelvin–Helmholtz contract to ignite off-centre O/Ne burning which result in the formation of convective region outside the flame. At the boundary of shell-flash convection, shear motion induced by convective flows and internal gravity waves lead to the mixing beyond the Schwarzschild convective boundaries, making the shell burning ashes intrude into the unburned material beneath the burning flame, which is known as convective boundary mixing. Previous works found that the convective boundary mixing could stall the inwardly propagating carbon flame in super-AGB stars and accreting CO WDs, resulting in the formation of CONe WDs or COSi WDs (e.g. Denissenkov et al. 2013; Farmer et al. 2015; Wu et al. 2020).

If a similar phenomenon occurs in the double ONe WD merger remnant, the death of the O/Ne flame would lead to qualitatively different results. In order to investigate the corresponding effect, we recalculate the ONe+ONe remnant by considering convective boundary mixing. In the ${\tt {MESA}}$ standard options, some parameterization of convective boundary mixing is generally employed alongside the mixing-length theory (MLT) to allow convective motions to extend outside of the MLT convection zone, and it allows us to take this mixing into account as a diffusion process. The corresponding diffusion coefficient in radiative layers adjacent to a convective boundary is (e.g. Freytag et al. 1996; Herwig 2000; Denissenkov et al. 2013)
\begin{equation}
    {D}_{\rm {CBM}}={D}_{\rm {MLT}}({r}_{\rm 0}){\rm {exp}}\left(-\frac{2|{r}-{r}_{\rm 0}|}{f{H}_{\rm p}}\right),
\end{equation}
where ${H}_{\rm P}$ is the pressure scale height and ${D}_{\rm {MLT}}({r}_{\rm 0})$ is a convective diffusion coefficient calculated based on MLT. The ${\tt {MESA}}$ ``mlt'' module assumes that ${D}_{\rm {MLT}}={\lambda}{v}_{\rm {conv}}/3$, where ${\lambda}={\alpha}{H}_{\rm P}$ is the mixing length (we use the default value of ${\alpha}=2$) and ${v}_{\rm {conv}}$ is the convective velocity. The radius ${r}_{\rm 0}$ is located at the distance of ${f}{H}_{\rm P}$ from the Schwarzschild boundary inside the convective zone. The free parameter ${f}$ should be calibrated through observations, or through hydrodynamic simulations. Similar to Farmer et al. (2015), we set ${f}$ equals to $0.016$ in this simulation. Meanwhile, in order to investigate the effects of electron capture and Urca cooling process, we reconstruct the $2.20{M}_{\odot}$ merger remnant with 5\% of $^{\rm 23}{\rm Na}$ and 1\% of $^{\rm 25}{\rm Mg}$ in the initial ONe WD (e.g. Schwab et al. 2017), and change our nuclear reaction network to ``${\tt {wd\_aic.net}}$'' in ``${\tt {MESA\,\,default}}$'' with artificially added isotopes (such as $^{\rm 20}{\rm O}$, $^{\rm 20}{\rm F}$, $^{\rm 24}{\rm Ne}$ and $^{\rm 20}{\rm Na}$) and the corresponding weak reactions. The weak interaction rates are from Schwab et al. (2015). We apply the ``${\tt {MESA\,\,default}}$'' screening factors to correct nuclear reaction rates for plasma interactions (e.g. Chugunov et al. 2007). The cooling rates from thermal neutrinos are derived from the fitting formulae of Itoh et al. (1996).

Fig.\,7 presents the structure of the merger remnant at the end of simulation under condition of convective boundary mixing. During the evolution, the convective flows penetrate the Schwarzschild convective boundaries, leading to the extra mixing between material at both sides of the flame, which decrease the O/Ne abundance ahead of the burning front. Since the nuclear reaction rate (${\epsilon}_{\rm nuc}$) is relevant to the abundance of fuel (${\epsilon}_{\rm nuc}$ can be approximated as ${\epsilon}_{\rm {nuc}}\propto{\rho}{\chi}^{\rm 2}{T}^{\rm n}$; i.e. nuclear reaction rate is more sensitive to the temperature, for example, for carbon burning at ${\rm {log}{T/{\rm K}}}=8.8-8.9$, ${n}\approx{\rm {40}}$), the O/Ne flame will quench gradually as the flame moves inward. Afterwards, the Kelvin-Helmholtz contract will increase the central density until it is high enough to trigger the electron capture reactions. We stop our calculation when the central temperature ${\rm {log}}{T/{\rm K}}>9.3$, where explosive O/Ne burning occurs. For the $2.20{M}_{\odot}$ merger remnant with overshooting parameter of ${f}=0.016$, the central O/Ne deflagration occurs when the inwardly propagating O/Ne flame reaches the mass coordinate of about $0.62{M}_{\odot}$. Eventually, the remnant will experience electron capture supernova to form an ONeFe WD \footnote{The mass of ONeFe WD is influenced by the resolution of simulations, the central density of ONe core and whether the Coulomb corrections are included in the equation of state during calculation. Hence, the mass of ONeFe WD would either be about $0.44-0.65{M}_{\odot}$ or $1.2-1.3{M}_{\odot}$ (see table\,1 of Jones et al. 2016 for detail).} due to its low central density (e.g. Jones et al. 2016, 2019).

\begin{figure*}
\begin{center}
\epsfig{file=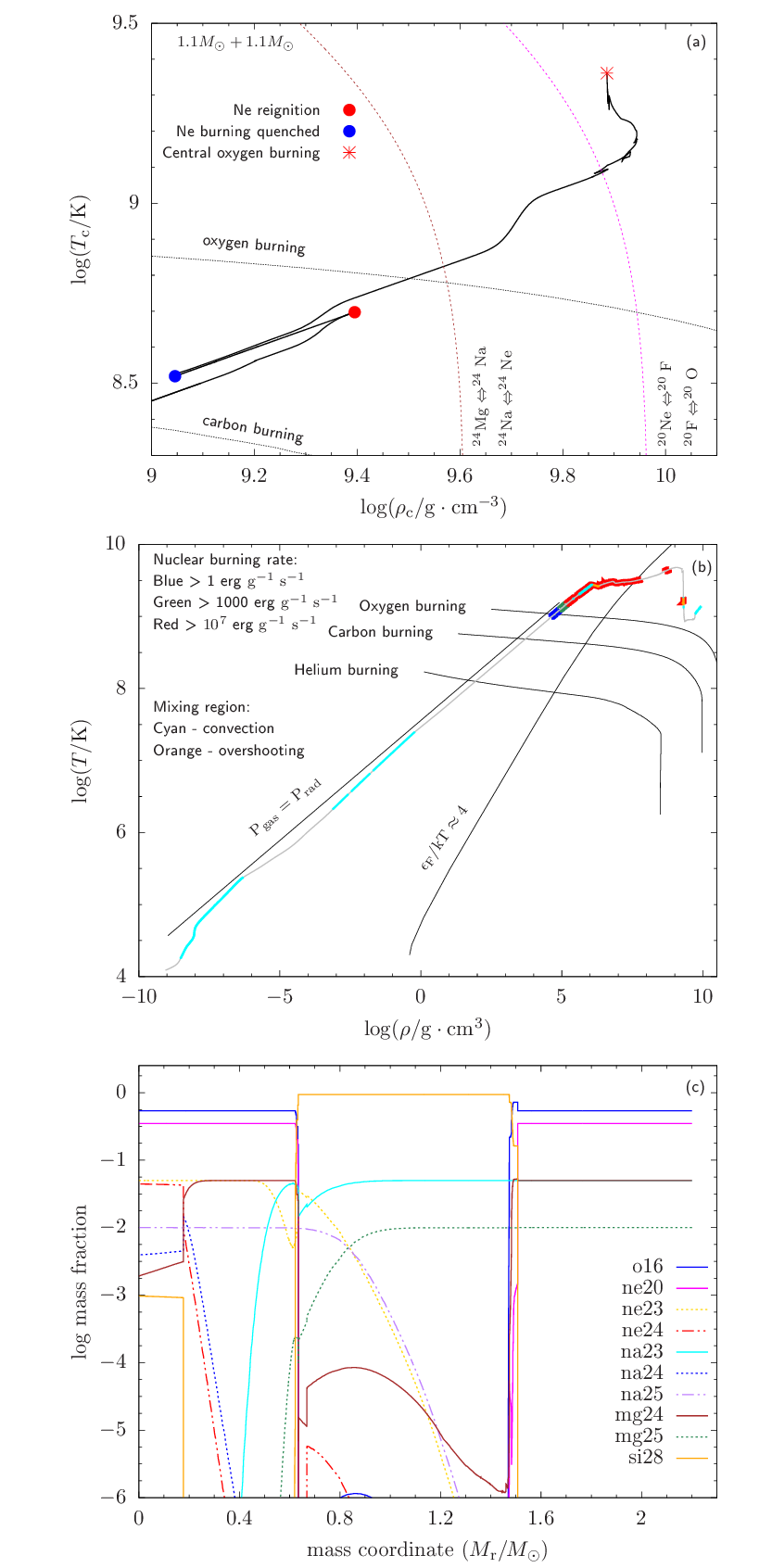,angle=0,width=10.2cm}
 \caption{Properties of the merger remnant when considering convective boundary mixing. Panel (a): the evolutionary track for central temperature-density. Black dotted lines represent carbon and oxygen burning lines, whereas brown and magenta dotted lines represent electron capture lines for $^{\rm 24}{\rm {Mg}}$ and $^{\rm 20}{\rm {Ne}}$. Panel(b): T-Rho profile for the merger remnant at the final stage when explosive O/Ne burning occurs. Black solid lines represent Pgas=Prad, dengerate line, helium, carbon and oxygen burning lines. Cyan line presents convective zone, while blue, green and red thick lines represent burning region with different nuclear reaction rates. Panel (c): the elemental abundance distributions of the merger remnant at its final evolutionary stage.}
  \end{center}
    \label{fig: 7}
\end{figure*}

\subsection{Wind mass-loss}

As the envelope responds to the energy deposited during the merger, the remnant begins to radiate away this energy which causes the remnant to inflate towards the giant structure. The stellar wind during the giant phase may be high enough to decrease the mass of the remnant and may further impact its final fate.

Since the envelope of the merger remnant is O/Ne dominated composition, it seems likely that these objects will drive dusty winds. The mass-loss rate of such objects is not observationally well constrained, and theoretically, there has no reliable way to estimate the corresponding mass-loss rate. Previous works usually assumed a wind mass-loss prescription to mimic the mass-loss rate of such merger remnant. For example, by assuming that the object will shed a fraction of its mass (${\beta}$) on a timescale $\frac{{\rm G}{M}^{\rm 2}}{{R}{L}}$ evaluating at the surface, Schwab (2021) suggested that the mass-loss rate of such remnant is
\begin{equation}
    {\dot{M}}={10}^{-5}{M}_{\odot}{\rm {yr}}^{-1}\left(\frac{\beta}{{10}^{-4}}\right)\left(\frac{R}{{100}{R}_{\odot}}\right)\left(\frac{L}{{3}\times{10}^{4}{L}_{\odot}}\right){\left(\frac{M}{{M}_{\odot}}\right)}^{-1}.
\end{equation}
In this case, they found that if ${\beta}={10}^{-5}$, the mass-loss rate for an $1.50{M}_{\odot}$ double CO WD merger remnant is equivalent to the one if we adopt Reimers' wind mass-loss prescription (e.g. Reimers 1975) with typical value for normal red giant of ${\eta}_{\rm {R}}=0.5$. Wu et al. (2023) assumed a Bl{\"o}cker wind mass-loss prescription (e.g. Bl{\"o}cker 1995) to investigated the effect of wind mass-loss process on ONe+CO WD merger remnants. They found that the wind mass-loss process could alter the final outcome for the merger remnants with masses lower than $1.95{M}_{\odot}$. Furthermore, they suggested that the $1.60{M}_{\odot}$ merger remnant under the Bl{\"o}cker's wind mass-loss rate with typical value for AGB stars (${\eta}_{\rm {B}}=0.05$) can explain the origin of CO-rich object - IRAS $00500+6713$.

In order to investigate the effect of the wind mass-loss process on double ONe WD merger remnant, we recalculate the $2.20{M}_{\odot}$ model by considering the Bl{\"o}cker's wind mass-loss prescription with ${\eta}_{\rm B}=0.05$. We evolve the remnant for $13.2$ years. At the time when we stop our simulation, the off-centre O/Ne flame reaches the position of ${M}_{\rm r}=0.018{M}_{\odot}$, and the remnant only shed ${\Delta}{M}\approx5\times{10}^{-5}{M}_{\odot}$. On the HR diagram, the evolutionary tracks of the merger remnant with and without wind mass-loss match, which means that the stellar wind process may not significantly impacting the final fate of double ONe WD merger remnants.

\subsection{Rotation}

The merger remnant could remain a portion of orbital angular momentum during the merger process. The condition of the O/Ne flame would be changed if the rotational velocity of the core is high enough, which may lead to a different evolutionary fate of the remnant. Previous works found that, for the double WD merger remnant, the angular momentum can be transported from the core to the envelope rapidly, and the core will have the angular velocity similar to the envelope several hours after merger. Afterwards, the angular velocity of the envelope will decrease obviously as the remnant inflate to the giant (e.g. Schwab 2021; Wu et al. 2023). In the present work, we have not simulated the evolution of rotational model. Nevertheless, according to the results of previous works, we infer that the rotation can hardly alter the evolution and final fate of double ONe WD merger remnant.

\section{Summary}

In this work, we simulated the evolution of double ONe+ONe WD merger remnant. The remnant underwent Kelvin–Helmholtz contract and ignited off-centre O/Ne burning soon after the merger. We found that it takes only about $20$ years for the fast moving O/Ne flame to reach the centre, making a short lifetime of the remnant. The final fate of the remnant is impacted by the process of convective boundary mixing. If the mixing process can prohibit the O/Ne flame from reaching the centre, the final outcome of the merger remnant would be an ONeFe WD through electron-capture supernova. Otherwise, the remnant may experience iron-core-collapse supernova to form an NS. Besides, due to the short lifetime of the remnant, we suggest that the wind mass-loss process may not affect the evolution and final fate of the remnant too much, unless an extremely strong wind mass-loss prescription is assumed. Finally, based on the results from previous works, we suggest that the rotation may not affect their evolution and final fate.

\section*{Acknowledgements}

We thank the referee Noam Soker for helpful comments. This study is supported by the National Natural Science Foundation of China (NSFC grants 12288102, 12003013, 12225304), the National Key R\&D Program of China (No. 2021YFA1600404), the Western Light Project of CAS (No. XBZG-ZDSYS-202117), the science research grant from the China Manned Space Project (No. CMS-CSST-2021-A12), the Yunnan Fundamental Research Projects (Nos 202301AU070039, 202001AS070029 and 202201BC070003),  the Frontier Scientific Research Program of Deep Space Exploration Laboratory (No. 2022-QYKYJH-ZYTS-016), and the International Centre of Supernovae, Yunnan Key Laboratory (No. 202302AN360001).

\section*{Data Availability}

Evolutionary models were computed with the version 12778 of MESA. The required inlists in this study can be available via request to the corresponding author.


\bsp	
\label{lastpage}
\end{document}